\documentclass{vak2024}
\usepackage[utf8]{inputenc}
\usepackage{url}
\usepackage{hyperref}

\begin{document}
\title{Exchange of meteorites between the terrestrial planets and the Moon}
\titlerunning{Exchange of meteorities}  % abbreviated title (for running head)
	%also used for the TOC unless
	%\toctitle is used
\author{Sergei~I.~Ipatov\inst{1}}
\authorrunning{Ipatov} % abbreviated author list (for running head)
    %\tocauthor{}
	%
\institute{$^{1}$ Vernadsky Institute of Geochemistry and Analytical Chemistry of RAS, Moscow, 119991, Russia}
\abstract{
 The evolution of the orbits of bodies ejected from the Earth, Moon, Mercury and Mars was studied.
The probabilities of collisions of ejected bodies with planets depended on ejection velocities, ejection angles and points of ejection.
At a velocity of ejection close to the parabolic velocity, most of bodies fell onto the planet from which they had been ejected. Below results are presented not for such small ejection velocities.
At ejection velocities about 12-14 km/s, the fraction of bodies ejected from the Earth that fall back onto the Earth was about 0.15-0.25. 
The total number of bodies ejected from the Earth and delivered to the Earth and Venus probably did not differ much. 
The probability of collisions of  bodies ejected from the Earth with the Moon moving in its present orbit was of the order of 0.01.
Probabilities of collisions of bodies ejected from the Earth with Mercury 
were about 0.02-0.08 at ejection velocities greater than 11.3 km/s. 
The probabilities of collisions of bodies ejected from the Earth with Mars 
did not exceed  0.025. 
For the ejection of bodies from the present orbit of the Moon, probabilities of collisions of ejected bodies with planets were similar to those ejected from the Earth if we consider smaller ejection velocities from the Moon than from the Earth.
 The probability of a collision of a body ejected from Mars with Mars usually did not exceed 0.04 at an ejection velocity greater than 5.3 km/s. The fraction of bodies ejected from Mars  and collided with Mercury was typically less than 0.08. Probabilities of collisions of bodies ejected from Mars with the Earth and Venus were about 0.1-0.2 (each) at an ejection velocity between 5.05 and 10 km/s. 
Most of bodies ejected from Mercury fall back onto Mercury. Probabilities of collisions of bodies ejected from Mercury with the Earth typically did not exceed 0.02 and 0.1 at 
an ejection velocity less than 8 km/s and 15 km/s, respectively. The fraction of bodies ejected from Mercury and collided with Venus was greater than that with the Earth typically by an order of magnitude. 
%The difference was greater for smaller ejection velocities.  
Probabilities of collisions of bodies with Venus  were about 0.1-0.3 at 
a velocity of ejection from Mercury between 4.3 and 10 km/s. 

\keywords{
%{Earth; Moon; 
planets and satellites: terrestrial planets; meteorites, meteors, meteoroids}
%meteorites; terrestrial planets; Moon; migration; ejection}
\doi{10.26119/VAK2024-143}
}

\maketitle

\section{Introduction and initial data}

 %Gladman et al.,  2005; Reyes-Ruiz M. et al., 2012) 
 \citet{5} and \citet{9}  studied the motion of bodies ejected from the Earth at collisions of bodies-impactors with the Earth  during  30 Kyr. Considered initial velocities were perpendicular to the surface of the Earth. 
%Wetherill (1984) 
 \citet{10} considered the evolution of the orbits of bodies ejected from Mars in random directions with velocity of ejection varied from 5 to 7 km/s. Gladman et al. (1995, 1996) and Gladman (1997) simulated the evolution of the orbits of bodies ejected from  the Moon (with ejection velocities of 2.3 and 3.5 km/s), Mars 
(with the values $v_{inf}=(v_{ej}^2-v_{esc}^2)^{1/2}$ of a velocity at the infinity from 1 to 3.3 km/s, where $v_{esc}$ is the escape velocity), and Mercury (with $v_{inf}$=1 km/s). They studied characteristic times before the collisions of these bodies with the Earth and the fraction of bodies that fell onto the Earth.  Gladman and Coffey (2009) studied the planetary-influenced motion of bodies ejected from Mercury with velocities equal to 4, 9, 14, 20, and 25 km/s. 

In each my calculation variant, the motion of $N_\circ$=250 bodies ejected from a celestial object was studied for fixed values of an ejection angle $i_{ej}$, a velocity $v_{ej}$ of ejection, a place of ejection,  and a time step $t_s$ of integration. In different variants, the values of $i_{ej}$ equaled to $15^\circ$, 30$^\circ$, 45$^\circ$, 60$^\circ$, $89^\circ$ or $90^\circ$. For ejection from the Earth $v_{ej}$ equaled mainly to 11.22, 11.5, 12, 12.7, 14, 16.4, or 20 km/s. In most calculations, bodies started directly from surface of a considered  celectial object. For Mars, Mercury, and Moon, $v_{ej}$ varied from a parabolic velocity on the surface of a celestial object to 20 km/s.
In each variant, bodies started from one of six considered opposite points on planet's surface (Ipatov, 2024).  
For the starting points F and C, the motion of the bodies started at most and least distant points of the planet's surface from the Sun (located on the line from the Sun to the planet), respectively. For the points W and B, the bodies started from the forward point on the planet's surface in the direction of the planet's motion and from the back point on the opposite side of the planet, respectively (from apex and antapex). For the points U and D, the bodies started from points on the planet's surface with the maximum and minimum values of z (with the z axis perpendicular to the plane of the planet's orbit), respectively. 

The motion of bodies was studied during the dynamical lifetime $T_{end}$ of all bodies, which equaled to a few hundred Myrs. During this time interval, all bodies collided with planets or the Sun or were ejected into hyperbolic orbits. The gravitational influence of the Sun and all eight planets was taken into account. Bodies that collided with planets or the Sun or reached 2000 AU from the Sun were excluded from integration. The symplectic code from the SWIFT integration package by  \citet{8}
was used for integration of the motion equations. The considered time integration step $t_s$ equaled to 1, 2, 5, or 10 days. For the Earth and Mars, the results of calculations with different $t_s$ were compared and gave similar results. For ejection from the Earth, Moon, and Mars most of calculations were done with $t_s$=5 days. Motion of bodies ejected from Mercury studied with $t_s$=1 day. Probabilities of collisions of bodies with the Moon were calculated based on the arrays of orbital elements of bodies (Ipatov, 2019, 2024).
Such ejection was often at the stage of accumulation of the terrestrial planets and at the Late Heavy Bombardment.

\section{Results of calculations}

\subsection{Probabilities of collisions of bodies ejected from the Earth with the Earth and the Moon} 
At ejection velocities $v_{ej}$$\le$11.25 km/s, i.e., slightly greater than the parabolic velocity, most of the  bodies ejected from the Earth fell back onto the Earth.
At $v_{ej}$ equal to 11.5 or 12 km/s, the values of the probabilities $p_e$ of collisions of bodies {\bf with the Earth} did not differ much for different starting points on the Earth's surface.
At $v_{ej}$=11.5 km/s, the probability $p_e$ of collisions of bodies ejected from the Earth with the Earth was about 0.2-0.3. 
At ejection velocities about 12-14 km/s, the value of $p_e$ 
%of bodies ejected from the Earth that fall back onto the Earth
 was about 0.15-0.25, mean velocities of collisions of bodies with the Earth and the Moon were about 14-20 and 10-16 km/s, respectively. The value of $p_e$ is greater for smaller ejection velocity. 
The ratio of the probability of collisions of bodies with the Earth to the probabilities of collisions of bodies with other planets and the Sun usually decreased with time. For example, the ratio of probabilities of collisions of bodies with the Sun at $T$=$T_{end}$ and at $T$=10 Myr could reach 5, although the ratio of $p_e$ at these times could be less than 1.5.
The significant difference in the results for different ejection points was  for bodies  ejected from the forward point  W of the Earth's motion at $v_{ej}$$\ge$$16.4$ km/s and 
$i_{ej}$$\ge$45$^\circ$. In this case, more than 80\% of the bodies were thrown into hyperbolic orbits, most of other bodies collided with the Sun, and $p_E$ could be 0.

The probability of collisions of  bodies ejected from the Earth {\bf with the Moon} moving in its present orbit was of the order of 0.01 (Ipatov, 2024). 
 A large Moon embryo should be formed close to the Earth in order to accumulate material ejected from the Earth's mantle which is not rich in iron. For more efficient growth of the Moon embryo, it is desirable that after some collisions of impactor bodies with the Earth, the ejected bodies did not simply fly out of the crater, but some of the matter went into orbits around the Earth, as in the multi-impact model.

\subsection{Probabilities of collisions of bodies ejected from the Earth with Venus, Mercury, Mars, and the Sun}
In considered calculations, the probabilities $p_v$ of collisions of the bodies ejected from the Earth  {\bf with Venus} at the end of evolution were often about 
0.2-0.35 at 11.5$\le$$ v_{ej}$$ \le$16.4 km/s. The values of $p_v$ were about $p_e$ at $v_{ej}$=11.5 km/s, were greater than $p_e$ at $v_{ej}$$\ge$12 km/s, and were less than $p_e$ at $v_{ej}$$\le$11.4 km/s. The total number of ejected bodies delivered to the Earth and Venus probably did not differ much. The obtained results testify in favour of that the upper layers of the Earth and Venus can contain similar material.
Probabilities of collisions of bodies ejected from the Earth {\bf  with Mercury} at $T$=$T_{end}$ were about 0.02-0.08 and 0.03-0.05 at 11.3$\le$$v_{ej}$$\le$11.5 and 12$\le $$v_{ej} $$\le$20 km/s, respectively. 
The probabilities of collisions of bodies {\bf with Mars} were smaller than those with Mercury and did not exceed 0.012 and 0.025 at $T$=10 Myr and $T$=$T_{end}$, respectively.
 More material ejected from the Earth was delivered to Mercury than to Mars. 
The probabilities of collisions of bodies with Jupiter were of the order of 0.001. 
The fraction of bodies colliding with the Sun often ranged from 0.2 to 0.5 at 11.3$\le$$v_{ej}$$\le$14 km/s. 
The fraction of bodies ejected into hyperbolic orbits was mainly greater for a greater ejection velocity. During a whole considered time interval it did not exceed 0.1 at $v_{ej}$$\le$12 km/s. At $v_{ej}$$\ge$16.4 km/s and  $i_{ej}$$\ge$60$^\circ$, all bodies were ejected from the Solar System if they started from the front (in the direction of the motion) point W of the Earth. At  $i_{ej}$=$45^\circ$ and $v_{ej}$=16.4 km/s, the fraction of bodies ejected from the Solar System was about 0.25-0.27 if points of ejection were not in the front or back of the Earth’s motion.    
The fractions of collisions of bodies ejected from the Earth with planets could be similar to the fractions of collisions of planetesimals with planets for planetesimals that were left in the feeding zone of the Earth at the late stages of its formation. More detailed studies of the migration of bodies ejected from the Earth are presented in the paper which is in press in Icarus.

\subsection{Probabilities of collisions of bodies ejected from the Moon with the Earth}
At the ejection of bodies from point F for the present orbit of the Moon (with its distance from the Earth $r_{ME}$=60$r_E$, where $r_E$ is the radius of the Earth), $T$=10 Myr, and 30$^\circ$$\le$$i_{ej}$$\le$$60^\circ$, $p_e$ was about 0.2–0.25 at $v_{ej}$=2.5 km/s, 0.13–0.14 at $v_{ej}$=5 km/s, and 0.06–0.07 at 12$\le$$v_{ej}$$\le$16.4 km/s. At $T$=$T_{end}$ and 15$^\circ$$\le$$i_{ej}$$\le$$89^\circ$  $p_e$ was about 0.27–0.35 at $v_{ej}$=2.5 km/s, 0.2-0.25 at $v_{ej}$=5 km/s, and 0.1-0.14 at 12$\le$$v_{ej}$$\le$16.4 km/s. That is, at velocities slightly greater than the parabolic velocity, the values of $p_e$ could be approximately the same for the ejection of bodies from the Earth and the Moon, but for different ejection velocities, 
 if we take into account the lower minimum velocities of bodies ejected from the Moon. 
For $v_{ej}$=11.22 km/s, $T$=10 Myr, 30$^\circ$$ \le $$i_{ej}$$\le$60$^\circ$, and $r_{ME}$=37$r_E$, the value of $p_E$ was about 0.04–0.13.
Bodies ejected from the lunar embryo, which was moving close to the Earth, fell back onto the Earth and the Moon if their ejection velocity was less than the corresponding parabolic velocity. 
At $v_{ej}$=2.5 km/s and $r_{ME}$=$3r_E$ the dynamical lifetime of the ejected bodies was less than 5 days. At $r_{ME}$=5$r_E$ and $v_{ej}$=2.5 km/s, most of the ejected bodies quickly fell onto the Earth, but at $i_{ej}$$\le$$30^\circ$ some of the bodies left the Hill sphere of the Earth.
At $v_{ej}$=5 km/s (approximately the parabolic velocity of the Earth at $r_{ME}$=5$r_E$), $T$=$T_{end}$, and $r_{ME}$=5$r_E$, the value of $p_e$ was about 0.3 (0.26-0.37) for $i_{ej}$ from 15$^\circ$ to 89$^\circ$.

\subsection{Probabilities of collisions of bodies ejected from Mars  with planets}
The probability $p_{ma}$ of a collision of a body ejected from Mars {\bf  with Mars} was considerable only at an ejection velocity $v_{ej}$ close to the parabolic velocity. Below the results
 for $v_{ej}$$\ge$$5.05$ km/s are presented. For such $v_{ej}$, the values of $p_{ma}$ were relatively small. For ejection of bodies from points  C, D, F, and U, the values of $p_{ma}$ were mainly about 0.04-0.25, 0.01-0.04, and 0-0.02 at 
5.05$\le$$v_{ej}$$\le$5.3, 5.5$\le$$v_{ej}$$\le$10, and
15$\le$$v_{ej}$$\le$20 km/s, respectively. For point B, the value of $p_{ma}$ was typically smaller than for the above points. 
For point W, $p_{ma}$=0 at 15$\le$$v_{ej}$$\le$20 km/s, and $p_{ma}$ could exceed 0.3 at $v_{ej}$=5.05 km/s. 
Probabilities $p_e$ of collisions of bodies ejected from Mars {\bf with the Earth} for points C, D, F, and U were mainly about 0.08-0.16 and 0-0.16 at 
5.05$\le$$v_{ej}$$\le$10 
and 15$\le$$v_{ej}$$\le$20 km/s, respectively. For point B, $p_e$ could exceed 0.24. 
For point W, the values of $p_e$ were in the range between 0 and 0.15. 
Often, at $v_{ej}$$\ge$5.2 km/s, the probabilities of collisions of bodies with Venus were slightly greater than those with the Earth.     
Probabilities $p_v$ of collisions of bodies ejected from Mars {\bf with Venus} for points C, D, F, and U were mainly about 0.08-0.2 and 0.02-0.2 
at 5.05$\le$$v_{ej}$$\le$10 and 15$\le$$v_{ej}$$\le$20 km/s, respectively. 
For point B, the value of $p_e$ could exceed 0.3. For point W, $p_e$ was in the range between 0 and 0.18.      
Bodies ejected from Mars could collide with the Earth after 0.1 Myr, and some of the bodies could collide with the Earth after hundreds of million years. For example, for point F at $v_{ej}$=6 km/s, $ i_{ej}$=45$^\circ$, $t_s$=5$^d$, and $N_\circ$=250, there were 32 collisions with the Earth between 0.22 and 270.9 Myr. Among these collisions there were 4, 7, 6, 7, 2, 4, and 2 collisions at time $t$$<$1, 1$<$$t$$<$5, 5$<$$t$$<$20, 20$<$$t$$<$50, 50$<$$t$$<$200, and $t$$>$200 Myr, respectively. In this case about a half of bodies collided with the Earth after 20 Myr, and some martian meteorites could travel for tens of million years in space before their collisions with the Earth. 
   The fraction of bodies ejected from  Mars and then {\bf colliding with Mercury} was usually less than 0.06 and was greater than the fraction of bodies colliding with Mars for $v_{ej}$$\ge$6 km/s.
For ejection from points C, D, F, and U, the values of the fraction $p_{me}$ of bodies collided with Mercury were mainly about 0.02-0.08 at 5.05$\le$$v_{ej}$$\le$20 km/s. For point B, the value of $p_{me}$ was in a wider range (0.016-0.2) than for the above four points. 
For point W, $p_{me}$=0 at 15$\le$$v_{ej}$$\le$20 km/s, and it could exceed 0.06 at $v_{ej}$=5.1 km/s. 
   The values of the fraction of bodies collided {\bf with the Sun} were typically between 0.2 and 0.9 for points C, D, F, U, and B. 
 Usually at 5.1$\le$$v_{ej}$$\le$8 km/s more than a half of bodies collided with the Sun. 
   The probability of ejection of a body into a hyperbolic orbit was less than 0.1 at $v_{ej}$$\le$6 km/s, but it could exceed 0.9 at $v_{ej}$=20 km/s.
For point W and 15$\le$$v_{ej}$$\le$20 km/s,  almost all bodies were ejected into hyperbolic orbits. For point B less than 10\% of bodies were ejected into hyperbolic orbits. For other four points, the probability of such ejection varied from 0.02 to 0.9 depending on an ejection velocity and an ejection angle.

\subsection{Probabilities of collisions of bodies ejected from  Mercury with planets}
Most of the bodies ejected from Mercury fell back onto Mercury. The probabilities of collisions of bodies ejected from Mercury with the Earth usually did not exceed 0.02 and 0.1 at $v_{ej}$$<$8 km/s and $v_{ej}$$<$15 km/s, respectively. The probabilities of their collisions with Venus were usually about 0.2-0.3 at 6$\le$$v_{ej}$$\le$10 km/s. The fraction of bodies colliding with Venus was usually greater by an order of magnitude than with Earth. The difference was larger at lower ejection velocities. The fraction of bodies ejected into hyperbolic orbits did not exceed 0.01 for most calculation variants.

%\acknowledgements{The studies were carried out under government-financed research project for the Vernadsky Institute.}

\section*{Funding} 
The studies were carried out under government-financed research project for  the Vernadsky  Institute of Geochemistry and Analytical Chemistry of the Russian Academy of Sciences.

%\bibliographystyle{aa}
%\bibliography{template}

\begin{thebibliography}{10}
\expandafter\ifx\csname natexlab\endcsname\relax\def\natexlab#1{#1}\fi

\bibitem[{Gladman, 1997}]{1}
{1. Gladman} B., 1997,  Icarus, Volume 130, pp. 228-246.

\bibitem[{Gladman and Coeffey, 2009}]{2}
{2. Gladman} B. and Coffey J., 2009,  Meteorit. Planet. Sci., Volume 44, pp. 285-291.

\bibitem[{Gladman et al., 1995}]{3}
{3. Gladman} B.J., Burns J.A., Duncan, M., et al.,
%Levison, H. F., 
1995, Icarus, Volume 118, pp. 302-321.

\bibitem[{Gladman et al., 1996}]{4}
{4. Gladman}, B.J., Burns, J.A., Duncan, M., et al.,
%Lee, P., Levison, H. F., 
1996, Science, Volume 271, pp. 1387-1392.

\bibitem[{Gladman et al., 2005}]{5}
{5. Gladman}, B., Dones, L., Levison H.F., et al.,
%Burns J.A.,  
2005. Astrobiology, Volume 5, pp. 483–496.

\bibitem[{Ipatov, 2019}]{6}
{6. Ipatov} S.I., 2019, Solar System Research, Volume 53, Issue 5, pp. 332-361. $https://doi.org10.1134/S0038094619050046$. $http://arxiv.org/abs/2003.11301$. 

\bibitem[{Ipatov, 2024}]{7}
{7. Ipatov} S.I., 2024,  Solar System Research, Volume 58, Issue 1, pp. 94-111. $https://doi.org/10.1134/S0038094624010040$.  $http://arxiv.org/abs/2405.19797$.

\bibitem[{Levison and Duncan} (1994)]{8}
{8. Levison}  H.F. and {Duncan}, M.J., 1994, Icarus, Volume 108, pp. 18–36. 

\bibitem[{Reyes-Ruiz et al.} (2012)]{9}
{9. Reyes-Ruiz} M., Chavez C.E., Aceves H., et al., 2012,  Icarus, Volume 220, pp. 777-786.

\bibitem[{Wetherill} (2012)]{10}
{10. Wetherill} G.W., 1984, Meteoritics, Volume 19, pp. 1-13.

 \end{thebibliography}

\end{document}